\documentclass[usenatbib,usegraphicx]{mn2e}
\usepackage{amsfonts}
\usepackage{amsmath}
\usepackage{amssymb}
\usepackage{comment}
\usepackage[usenames,dvipsnames]{color}

\title[Radio AGN in spiral galaxies]
    {Radio AGN in spiral galaxies}
\author[Sugata Kaviraj]
{Sugata Kaviraj\thanks{s.kaviraj@herts.ac.uk}$^{1}$, Stanislav S.
Shabala$^{2}$, Adam T. Deller$^{3}$ and Enno Middelberg$^{4}$\\
$^{1}$Centre for Astrophysics Research, University of
Hertfordshire, College Lane, Hatfield, Herts, AL10 9AB, UK\\
$^{2}$School of Mathematics and Physics, University of Tasmania, Private Bag 37, Hobart, TAS 7001, Australia\\
$^{3}$The Netherlands Institute for Radio Astronomy (ASTRON),
Dwingeloo, The Netherlands\\
$^{4}$Astronomisches Institut der Ruhr-Universit\"{a}t Bochum,
Universit\"{a}tsstra\ss e 150, D-44801 Bochum, Germany}

\begin{document}

\maketitle

\def \aj {AJ}
\def \mnras {MNRAS}
\def \pasp {PASP}
\def \apj {ApJ}
\def \apjs {ApJS}
\def \apjl {ApJL}
\def \aap {A\&A}
\def \nat {Nature}
\def \araa {ARAA}
\def \iaucirc {IAUC}
\def \aaps {A\&A Suppl.}
\def \qjras {QJRAS}
\def \na {New Astronomy}
\def \aapr {A\&ARv}
\def\lesssim{\mathrel{\hbox{\rlap{\hbox{\lower4pt\hbox{$\sim$}}}\hbox{$<$}}}}
\def\gtrsim{\mathrel{\hbox{\rlap{\hbox{\lower4pt\hbox{$\sim$}}}\hbox{$>$}}}}


\begin{abstract}
Radio AGN in the nearby Universe are more likely to be found in
galaxies with early-type morphology, the detection rate in spiral
or late-type galaxies (LTGs) being around an order of magnitude
lower. We combine the mJy Imaging VLBA Exploration at 20cm
(mJIVE-20) survey with the Sloan Digital Sky Survey (SDSS), to
study the relatively rare population of AGN in LTGs that have
nuclear radio luminosities similar to that in their early-type
counterparts. The LTG AGN population is preferentially hosted by
galaxies that have high stellar masses (M$_* > 10^{10.8}$
M$_{\odot}$), red colours and low star-formation rates, with
little dependence on the detailed morphology or local environment
of the host LTG. The merger fraction in the LTG AGN is $\sim$4
times higher than that in the general LTG population, indicating
that merging is an important trigger for radio AGN in these
systems. The red colours of our systems extend recent work which
indicates that merger-triggered AGN in the nearby Universe appear
\emph{after} the peak of the associated starburst, implying that
they do not strongly regulate star formation. Finally, we find
that in systems where parsec-scale jets are clearly observed in
our VLBI images, the jets are perpendicular to the major axis of
the galaxy, indicating strong alignment between the accretion disc
and the host galaxy stellar disc.
\end{abstract}




\section{Introduction}
Radio AGN are a cornerstone of modern galaxy formation models, the
current consensus favouring a paradigm in which AGN feedback
regulates star formation and shapes fundamental features of the
galaxy population like their distribution of colours and
luminosities \citep[e.g.][]{Hatton2003,Croton2006,Somerville2012}.
Observational work indicates that radio AGN, at least in the local
Universe, are more common in early-type galaxies, the detection
rate in spiral or late-type galaxies (LTGs) being around an order
of magnitude lower \citep{Ledlow2001,Kaviraj2014a}. Radio AGN are,
therefore, comparatively rare in LTGs compared to their early-type
counterparts, with relatively few examples having been studied in
the past literature
\citep[e.g.][]{Ekers1978,Ulvestad1984,Edelson1987,Ledlow1998,Morganti2001,Keel2006,Zhou2007,Yuan2008,Gliozzi2010,Inskip2010,Norris2012,Mao2014}.

\begin{figure*}
\begin{minipage}{172mm}
\begin{center}
$\begin{array}{c}
\includegraphics[width=\textwidth]{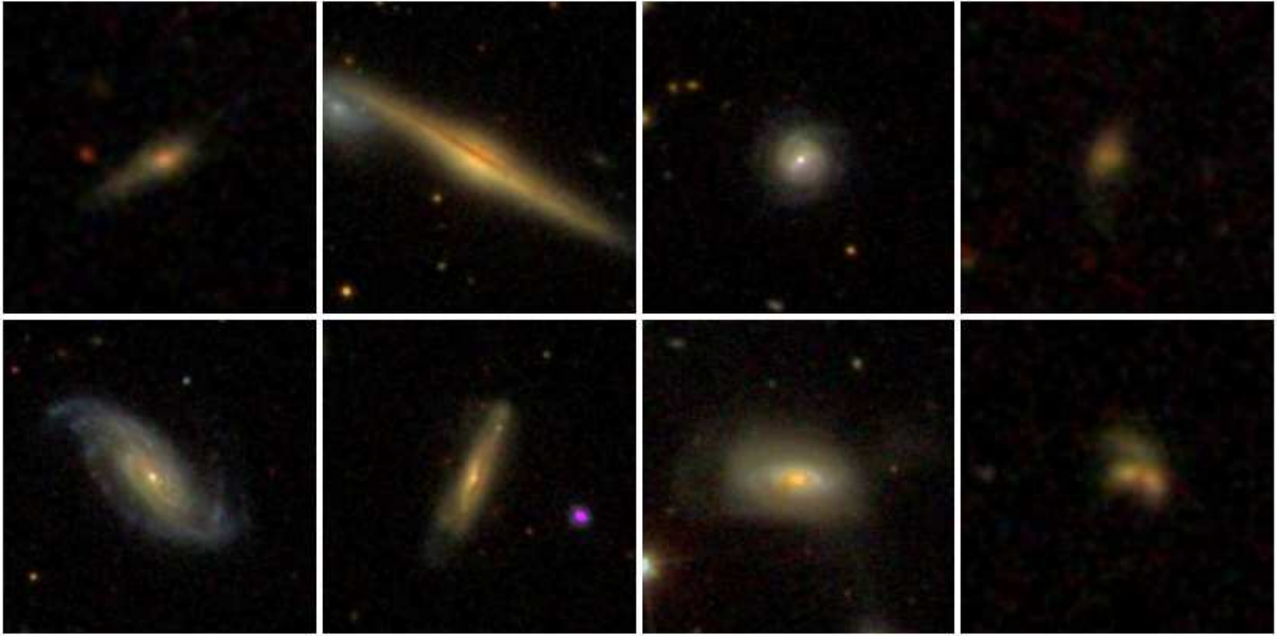}
\end{array}$
\caption{SDSS colour images of a sample of VLBI-detected spiral
galaxies that underpin this study.} \label{fig:spirals}
\end{center}
\end{minipage}
\end{figure*}

The study of radio AGN in LTGs is desirable not just because they
are rare, but also because they may hold clues to the behaviour of
radio jets in disc environments. While this is uncommon at low
redshift, the bulk of the stellar and black-hole mass in today's
Universe was created at $z\sim2$
\citep[e.g.][]{Madau1998,Hopkins2006}, an epoch at which both star
formation \citep[e.g.][]{Kaviraj2013a} and black-hole growth
\cite[e.g.][]{Kocevski2012,Schawinski2012} were predominantly
hosted by late-type galaxies. The connection between the black
hole and a disc-like host system was therefore common around the
epoch of peak cosmic star formation, making radio AGN in nearby
LTGs useful laboratories for exploring this connection.

Given their relative rarity, a statistical study of radio AGN in
LTGs requires a combination of survey-scale radio data (for the
detection of large numbers of radio AGN) and high-resolution
optical imaging (for accurate determination of galaxy
morphologies). In this study, we combine mJIVE-20 (Deller \&
Middelberg 2014) with optical imaging and spectro-photometry from
the SDSS (Azabajian et al. 2009), to study radio AGN in the nearby
($z<0.3$) LTG population.

The high resolution of VLBI enables unambiguous identification of
AGN, because the $>10^6$ K temperatures required for a detection
cannot be achieved via star formation alone and requires
non-thermal sources like supernova remnants (SNRs), radio
supernovae (SNe) or AGN. Only in the very local Universe and in
vigorous starbursting systems can clusters of luminous SNe and
SNRs reach sufficiently high luminosities to be visible in a
(shallow) VLBI observation. In the extreme example of Arp 220,
which has a star formation rate of several hundred solar masses
per year \citep[e.g.][]{Iwasawa2005,Baan2007}, the brightest
VLBI-scale sources reach a peak VLBI flux density of $\sim$1
mJy/beam \citep{Lonsdale2006}. As we indicate below, our sample
consists of sources that are both more distant and have lower star
formation rates than Arp 220, making it extremely unlikely that
the relatively shallow mJIVE-20 observations (with a detection
limit of $\sim$1 mJy) would detect anything other than an AGN in
our target sample.

While in lower-resolution surveys like FIRST and NVSS, AGN
identification requires the detection of excess radio flux beyond
what is expected from star formation (biasing samples towards AGN
that dominate the star formation), or a dense gas environment
which the jet works against to produce detectable radio lobes
(introducing a bias against AGN in LTGs, which do not typically
host extensive hot gas halos), VLBI is capable of identifying AGN
irrespective of the host galaxy properties (e.g. star formation
rate, environment etc.). Some limitations of VLBI are worth noting
here. Of the total radio emission generated by AGN activity, only
a fraction will typically be confined to the parsec-scale core/jet
at the site of the central black hole, as extended radio lobes
will often be present -- hence, the compact flux fraction can be
influenced by the surrounding galactic environment. The prominence
of the compact core depends on the source age and orientation
(which can result in Doppler boosting or deboosting of the compact
emission). Finally, hotspots at the site of the jet interaction
with the interstellar medium \citep[as are seen in compact
symmetric objects -- CSOs][]{Phillips1982,Wilkinson1994} may also
be compact enough to be visible in VLBI observations, so that,
although VLBI detections can unambiguously be associated with AGN,
they cannot in every case be associated with AGN {\em cores}.
However, this last case can usually be distinguished on the basis
of morphology. Nevertheless, the unambiguity in the identification
of AGN activity makes VLBI data from mJIVE-20 the ideal route for
studying the local AGN population that inhabits LTGs.

Here, we identify our spiral AGN hosts via direct visual
inspection of the SDSS colour images of mJIVE-20 detections. We
then study their physical properties, derived from SDSS
spectro-photometric data, to explore the conditions that make it
likely for LTGs to host AGN. This is the first study where
survey-scale VLBI data is combined with visual inspection of
galaxy images to study a \emph{morphologically-selected} sample of
LTGs that host radio AGN. The sections below are organized as
follows. In Section 2, we briefly describe the mJIVE-20 and SDSS
data that underpin this study. In Section 3, we compare the
properties of LTGs that host radio AGN with those of the general
LTG population. We discuss and summarize our findings in Section
4. Throughout, we employ the WMAP7 cosmological parameters
\citep{Komatsu2011} and photometry in the AB system
\citep{Oke1983}.


\section{Data}

\subsection{Radio VLBI data}
mJIVE-20 is an ongoing survey using the VLBA that is
systematically observing objects detected by the FIRST radio
survey. mJIVE-20 uses short segments scheduled in bad weather or
with a reduced number of antennas during which no highly rated
VLBA science projects can be scheduled. The survey has targeted
$\sim$25,000 FIRST sources to date, with $\sim$5000 VLBI
detections. While the sensitivity and resolution of mJIVE-20
varies between different fields, the median detection threshold is
1.2 mJy/beam and the typical beam size is 6$\times$17
milliarcseconds. This corresponds to a detection sensitivity of
$\sim10^7$ K (the variation between fields is around a factor of
2). We refer readers to \citet{Deller2014} for further details of
the survey.


\subsection{SDSS data}
The mJIVE-20 targets are cross-matched with the latest data
release of the Sloan Digital Sky Survey
\citep[SDSS;][]{Abazajian2009}. Following \citet{Shabala2008}, we
use a matching radius of 2 arcseconds for the radio-optical
matching, which yields high (96\%) completeness and low (0.3\%)
contamination. Since morphological classification and the
identification of morphological disturbances will be an important
part of our study, we restrict ourselves to SDSS galaxies that
have photometric or spectroscopic redshifts less than 0.3. This
yields 437 SDSS galaxies with mJIVE-20 detections in this redshift
range. The entire sample is visually classified, via the SDSS
colour images, to separate early-type galaxies from LTGs. For each
galaxy, we also note the presence of morphological disturbances,
indicating that the galaxy has had a recent merger or interaction.

This visual classification yields 29 VLBI-detected LTGs. Figure
\ref{fig:spirals} presents examples of these systems. Note that
the VLBI-detected LTGs span the full spectrum of spiral
morphologies seen on the Hubble sequence
\cite[e.g.][]{Hubble1926,dev1959}, from systems that have
reasonably prominent bulges to those that are clearly dominated by
discs. It is also worth noting that the VLBI-detected LTGs show a
high incidence of morphological disturbances, suggesting that
these systems have undergone recent mergers (we return to this
point in our analysis below).

Out of the 29 VLBI-detected LTGs, 11 have SDSS spectra and
therefore spectroscopic redshifts. Our analysis below is
restricted to these objects, since quantities like absolute
magnitudes, stellar masses and local environments require an
accurate measurement of redshift. For this
spectroscopically-detected subsample, magnitudes are K-corrected
using the \texttt{KCORRECT} code of \citet{Blanton2007} and
published stellar masses, star formation rates and emission-line
classes are extracted from the latest version of the
publicly-available MPA-JHU value-added SDSS catalogue
\citep[][]{Kauffmann2003,Brinchmann2004,Tremonti2004}\footnote{http://www.mpa-garching.mpg.de/SDSS/DR7/}.

The emission-line class of each galaxy has been calculated through
a standard line-ratio analysis \citep[][see also Baldwin et al.
1981, Veilleux et al. 1987, Kewley et al. 2006]{Kauffmann2003},
using the measured values of [NII]/H$\alpha$ and [OIII]/H$\beta$.
Objects in which all four emission lines are detected with a
signal-to-noise (S/N) ratio greater than 3 are classified in the
MPA-JHU catalog as either `star-forming', `low S/N star-forming',
`composite', `Seyfert' or `LINER', depending on their location in
the [NII]/H$\alpha$ vs. [OIII]/H$\beta$ diagram. Galaxies without
a detection in all four lines are classified as `quiescent'
\citep{Kauffmann2003}.

Finally, the local environments of these
spectroscopically-detected galaxies are extracted from the group
catalogue of \citet{Yang2007}, who use a halo-based group finder
to separate the SDSS into over 300,000 structures
{\color{black}with a broad dynamic range}, {\color{black}from}
rich clusters to isolated galaxies. \citet{Yang2007} estimate the
host dark matter (DM) halo masses of {\color{black}individual}
SDSS galaxies, which are related to the traditional
classifications of environment (`field', {\color{black}`group'}
and {\color{black}`cluster'}). Cluster-sized haloes typically have
masses greater than $10^{14}$ M$_{\odot}$, while group-sized
haloes have masses between $10^{13}$ and $10^{14}$ M$_{\odot}$
\citep{Binney1987}.


\section{Radio AGN in spiral galaxies}
Given that VLBI detections are possible in rare,
intensely-starbursting galaxies like Arp 220, we first check
whether our spiral AGN hosts are consistent with an Arp 220-like
starburst. The VLBI-detected LTGs have SFRs (Figure 5) that are
significantly lower than that in Arp 220, which has an SFR of
several hundred solar masses per year
\citep[e.g.][]{Iwasawa2005,Baan2007}. They also lie much further
away than Arp 220, which has a redshift of 0.018 (Figure 3). The
VLBI radio fluxes in these systems are, therefore, inconsistent
with being driven by star formation in an Arp 220-like system.
This conclusion is consistent with the SDSS images of these
galaxies (see Figure 1), which indicate that most of these objects
are not dusty major-merger remnants like Arp 220, but are normal
massive galaxies with a disc-like morphology.

\begin{figure}
$\begin{array}{c}
\includegraphics[width=3.5in]{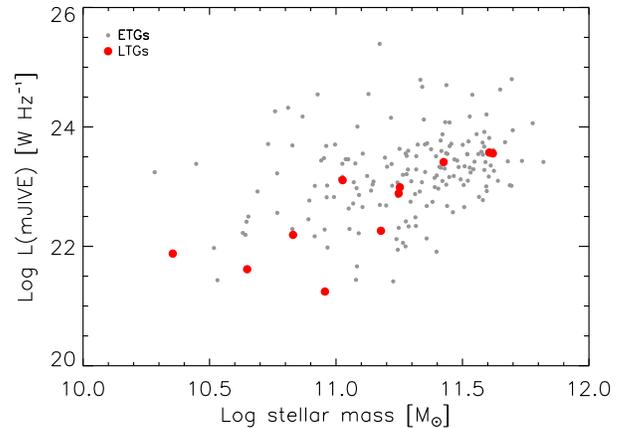}
\end{array}$
\caption{mJIVE-20 radio luminosities of VLBI-detected spirals
compared to that of their early-type counterparts in the same
stellar mass and redshift range.}
\label{fig:mJIVE-20_luminosities}
\end{figure}


\subsection{Physical properties of spiral AGN hosts}
We begin our analysis by studying the physical properties of the
LTGs that host radio AGN. Figure \ref{fig:mJIVE-20_luminosities}
indicates that the radio AGN in our LTG sample have similar masses
and nuclear radio luminosities as their early-type counterparts.
We are, therefore, studying radio AGN in spirals that broadly
share the same physical properties as the more commonly-found
radio AGN in early-type systems in the same stellar mass and
redshift ranges.

\begin{table*}
\begin{center}
\caption{Comparison of the properties of VLBI-detected,
VLBI-undetected and control LTGs. Columns are as follows: (1)
Galaxy number fraction on the red sequence, defined as $u-r>2.2$
(2) The fraction of galaxies that are currently merging or are
post-mergers (3-8) Fraction of galaxies in each BPT classification
class (star-forming [SF]), low S/N star forming, composites [Cp],
Seyferts [Sy], LINERs [LI] and quiescent [Qs]).}
\begin{tabular}{ccccccccc}\hline
                         & Red             & Mgr             & SF              & Low S/N SF      & Cp              & Sy              & LI              & Qs \\\hline
    VLBI-detected LTGs   & 0.90$^{(0.35)}$ & 0.53$^{(0.21)}$ & 0.08$^{(0.07)}$ & 0.08$^{(0.07)}$ & 0.23$^{(0.17)}$ & 0.30$^{(0.18)}$ & 0.23$^{(0.17)}$ & 0.08$^{(0.07)}$\\
    VLBI-undetected LTGs & 0.48$^{(0.16)}$ & 0.28$^{(0.06)}$ & 0.14$^{(0.04)}$ & 0.07$^{(0.03)}$ & 0.41$^{(0.08)}$ & 0.29$^{(0.06)}$ & 0.05$^{(0.02)}$ & 0.02$^{(0.01)}$\\
    Control LTGs         & 0.62$^{(0.06)}$ & 0.16$^{(0.02)}$ & 0.12$^{(0.02)}$ & 0.41$^{(0.04)}$ & 0.12$^{(0.02)}$ & 0.06$^{(0.02)}$ & 0.19$^{(0.03)}$ & 0.10$^{(0.02)}$\\

\end{tabular}
\end{center}
\label{tab:merger_fractions}
\end{table*}

\begin{figure}
$\begin{array}{c}
\includegraphics[width=3.5in]{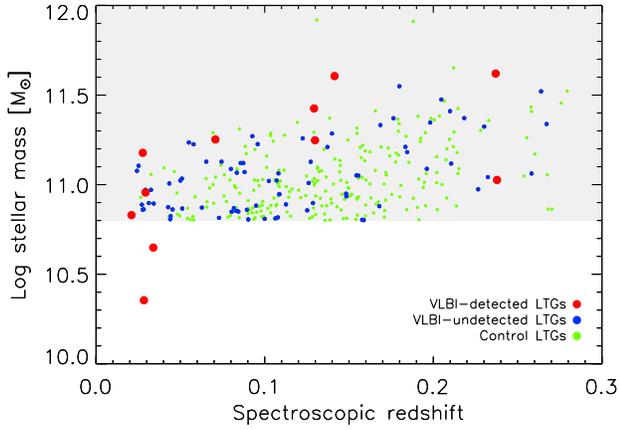}\\
\end{array}$
\caption{Stellar mass vs redshift of VLBI-detected (red) and
VLBI-undetected (blue) and control (green) spiral galaxies. The
shaded region indicates the stellar mass range considered in our
analysis in Section 3.} \label{fig:mass_redshift}
\end{figure}

\begin{figure}
$\begin{array}{c}
\includegraphics[width=3.5in]{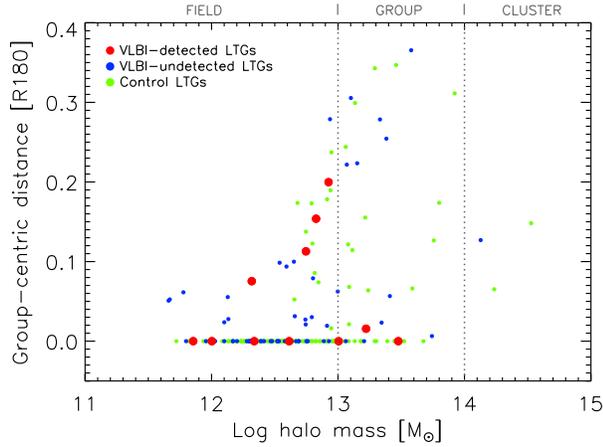}\\
\end{array}$
\caption{Halo mass vs group-centric radii of VLBI-detected (red),
VLBI-undetected (blue) and control (green) spiral galaxies.}
\label{fig:environment}
\end{figure}

\begin{figure}
$\begin{array}{c}
\includegraphics[width=3.5in]{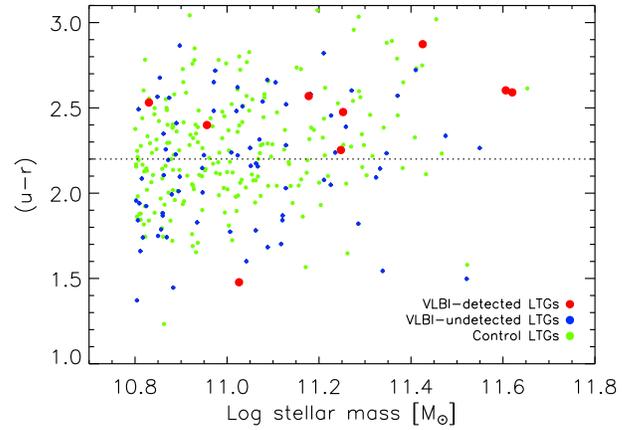}\\
\includegraphics[width=3.5in]{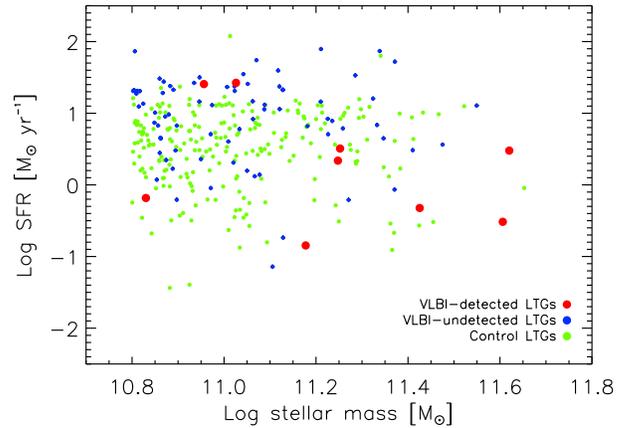}
\end{array}$
\caption{$(u-r)$ (top) and SFR (bottom) vs stellar mass for the
VLBI-detected (red), VLBI-undetected (blue) and control (green)
spiral galaxies. The dotted line in the top panel indicates the
demarcation between the red sequence and the blue cloud
\citep[e.g.][]{Strateva2001}.} \label{fig:sfr}
\end{figure}

In our subsequent analysis, we compare the physical properties of
our VLBI-detected LTGs to two different populations of LTGs. Since
most of our VLBI-detected LTGs have stellar masses greater than
M$_* \sim 10^{10.8} M_{\odot}$ (see Figure
\ref{fig:mass_redshift}) and have been restricted to $z<0.3$, we
restrict all other samples to these stellar mass and redshift
ranges. The first population is the sample of 77 LTGs that are
mJIVE-20 targets but are undetected (we refer to these galaxies as
`VLBI-undetected LTGs'). Since the VLBI-undetected LTGs are, by
construction, detected by FIRST, we also define a `control sample'
of LTGs that are not radio detected. To construct this control
sample we select a random $\sim$230 spiral galaxies from the SDSS
via an identical visual inspection as was performed on the
mJIVE-20 targets. Figure \ref{fig:mass_redshift} presents the
stellar mass and redshift properties of each of these LTG
populations. It is worth noting that the VLBI-detected spirals
generally lie towards the upper envelope of the mass-redshift
space. KS-tests between the mass distributions of the
VLBI-detected LTGs compared to the VLBI-undetected and control
LTGs yield p-values of $\sim$10$^{-5}$ and $\sim$10$^{-4}$
respectively, indicating that the spirals with AGN are likely to
be drawn from different parent distributions. The offset towards
higher masses in the VLBI-detected systems is plausibly due to
larger galaxies having bigger black holes \citep[][]{Gultekin2009}
which, in turn, produce more powerful jets
\citep[e.g.][]{Shabala2008}.

\begin{figure}
$\begin{array}{c}
\includegraphics[width=3.5in]{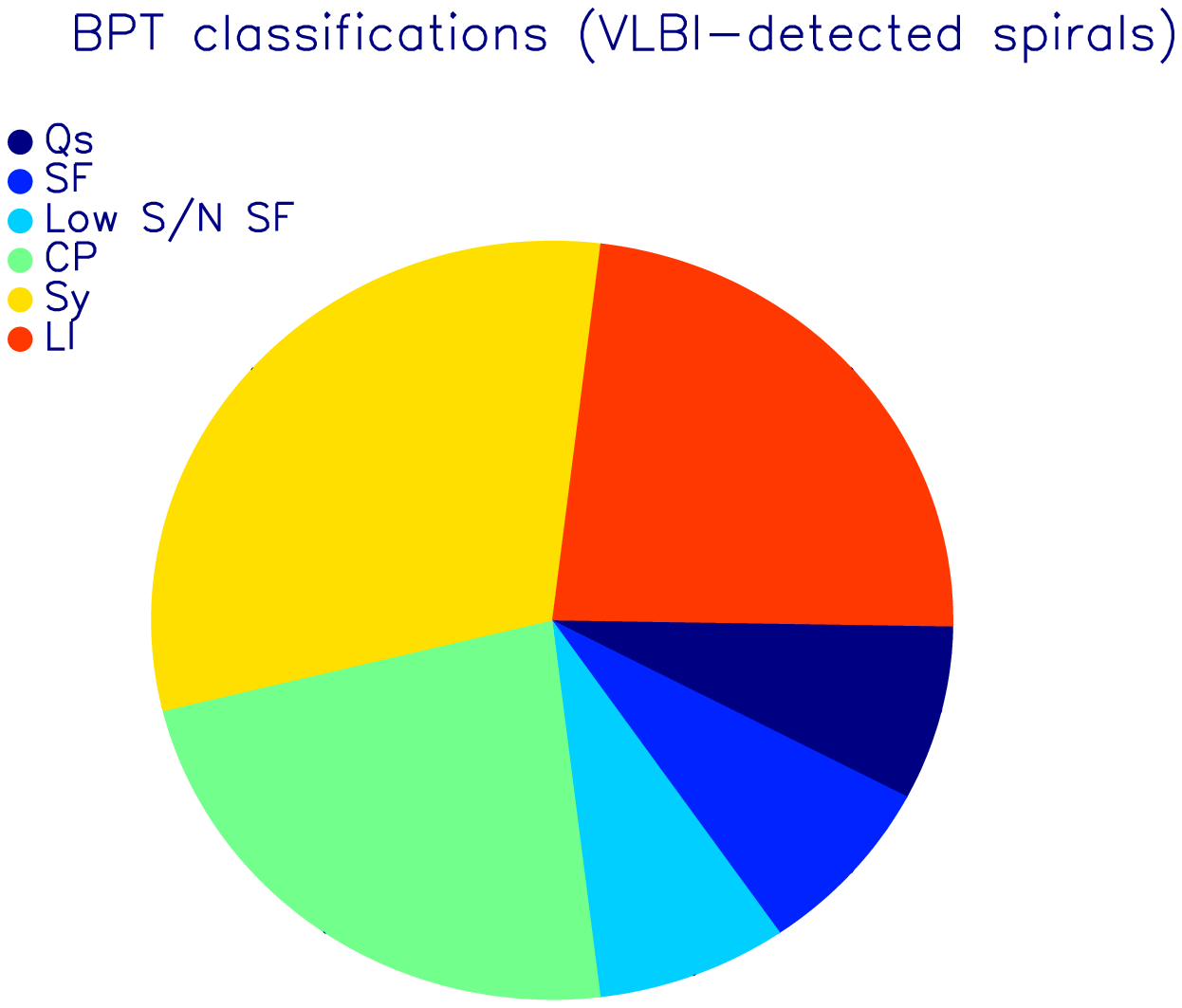}\\
\includegraphics[width=3.5in]{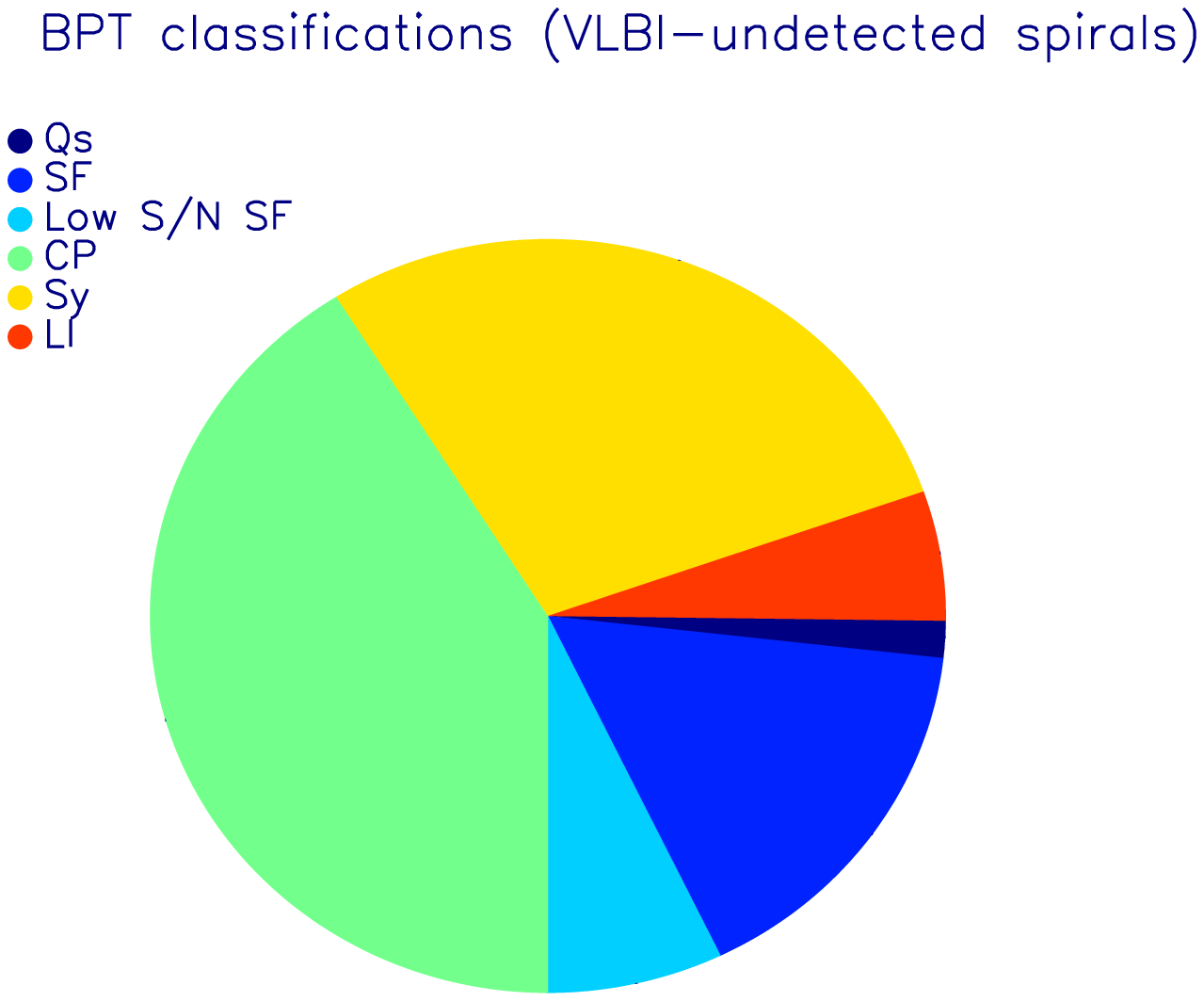}\\
\includegraphics[width=3.5in]{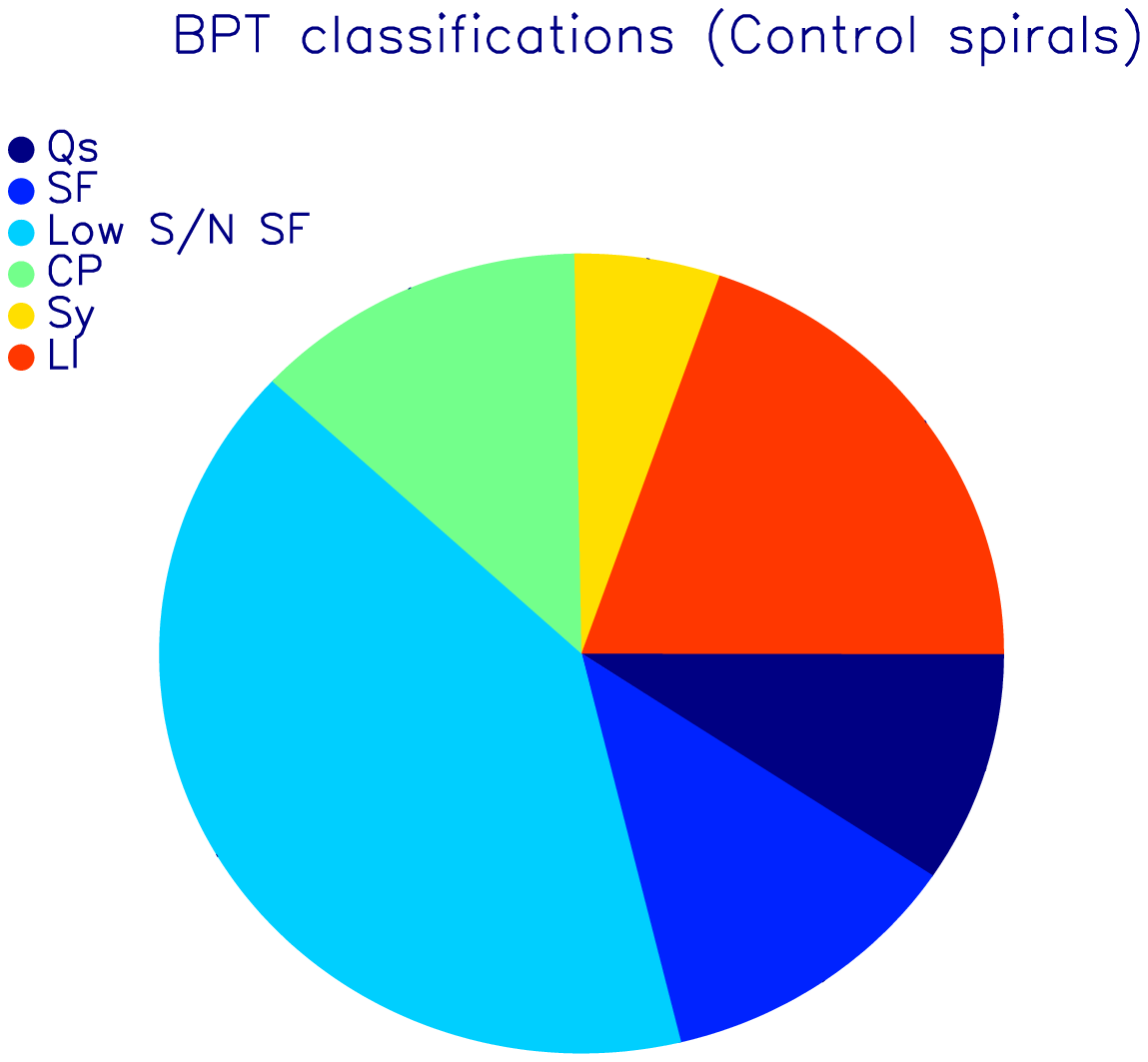}
\end{array}$
\caption{BPT classifications of VLBI-detected (top),
VLBI-undetected (middle) and control (bottom) spiral galaxies. The
fractions of galaxies in each morphological class are indicated in
Table 1.} \label{fig:bpt}
\end{figure}

Figure \ref{fig:environment} shows the environments of the spiral
AGN hosts compared to that of their VLBI-undetected and control
counterparts. Paired (i.e. 2D) KS-tests in the halo mass - group
centric radius plane between the VLBI-detected LTGs and their
VLBI-undetected and control counterparts yield p-values of 0.06
and 0.01 respectively. Thus, while the environments of the
VLBI-detected and undetected populations are likely to be similar,
there is a hint that the VLBI-detected LTGs are somewhat different
from the control LTGs. The difference is probably driven by the
fact that the VLBI detections appear to avoid the the highest halo
masses compared to the control sample. However, local environment,
on the whole, appears not to play a major role in determining the
presence of radio AGN in LTGs.

The VLBI-detected spirals generally show redder $(u-r)$ colours
and correspondingly lower SFRs than the undetected and control
populations (Figure \ref{fig:sfr}). KS-tests between the
VLBI-detected and control LTGs in either of these quantities yield
very low p-values ($<$10$^{-3}$), indicating that the spiral AGN
hosts are likely to be different from the general LTG population
in these properties. $\sim$90\% of the VLBI-detected LTGs are on
the red sequence, compared to $\sim$50\% of the VLBI-detected
sample and $\sim$60\% of the control LTG sample (column 1 in Table
1). In a similar vein, only 2 out of the 9 VLBI-detected LTGs lie
on the SFR-mass parameter space defined by the bulk of the control
and VLBI-undetected populations \footnote{It is worth noting here
that some of our spiral AGN are on the star formation main
sequence, illustrating the utility of VLBI in being able to detect
AGN regardless of the level of star formation activity in the host
system.}.

As has been noted in the literature \citep[e.g.][]{Greene2013},
the mechanism that dissipates angular momentum and allows gas to
accrete on to the central black hole remains unclear. The removal
of angular momentum could take place as a result of processes such
as merging \citep[e.g.][]{Sanders1988}, or via circumnuclear
structures such as bars \citep[e.g.][]{Maciejewski2002,Kim2012} or
nuclear spirals
\citep[e.g.][]{Englmaier2000,Martini2003,Davies2009}. Column 2 in
Table 2 presents merger fractions in our three LTG samples. The
VLBI-detected LTGs show the highest merger fraction ($\sim$50\%),
which is around a factor of 2 higher than that in their undetected
counterparts and around a factor of 4 higher than that in the
control LTGs. While other processes that drive the dissipation of
angular momentum cannot be completely ruled out, the remarkably
high merger fraction in the VLBI-detected spirals suggests that
mergers are likely to play a significant role in triggering radio
AGN in spiral hosts. It is worth noting that the SDSS images that
form the basis of the visual inspection and the identification of
morphological disturbances are relatively shallow, with standard
exposure times of 54 seconds. As noted in the recent literature
\citep[e.g.][]{Kaviraj2010}, merger fractions derived from such
images are, therefore, strictly lower limits. It is possible (and
likely) that a much larger fraction of the VLBI-detected LTGs
carry morphological disturbances, that are invisible in the
shallow standard-depth SDSS images.

A merger trigger for the AGN appears to be supported by the
emission-line analysis of our sample (Figure \ref{fig:bpt} and
columns 4-8 in Table 2). The BPT analysis suggests an increasing
fraction of optical emission-line AGN as we transition from the
control LTG sample to the VLBI-detected population. While
$\sim$37\% of the control sample show evidence for an optical AGN
(i.e. systems classified as composite, Seyferts or LINERs), the
corresponding fraction in the VLBI-detected population is
$\sim$76\%. It is worth noting that in early-type galaxies, some
LINERs may not be driven by AGN \citep[e.g.][]{Sarzi2010}. Even if
this was also the case in LTGs and the LINERs were removed from
this argument, the VLBI-detected spirals still show a surfeit of
optical AGN of around a factor of 3. This high fraction of optical
AGN appears consistent with a radiatively-efficient cold-mode type
accretion, which might be expected if the gas has been accreted
via a merger, as seems to be the case for our sample
\citep[e.g.][]{Hardcastle2007,Best2012,Shabala2012}. Our results
appear consistent with recent work that has suggested a strong
connection between merging and the onset of AGN activity at low
redshift
\citep[e.g.][]{Koss2010,Ellison2011,Scott2014,Kaviraj2014a},
although the role of mergers in triggering AGN may become
negligible at high redshift
\citep[e.g.][]{Chiaberge2011,Schawinski2012,Kocevski2012}.

\begin{figure*}
\begin{minipage}{172mm}
\begin{center}
$\begin{array}{ccc}
\includegraphics[width=2.1in]{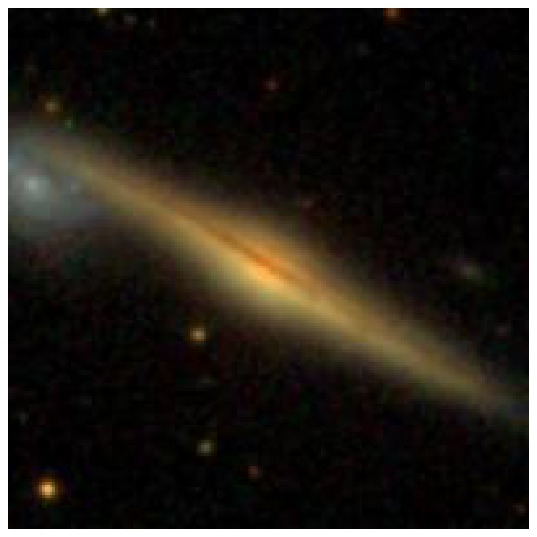}  & \includegraphics[width=2.1in]{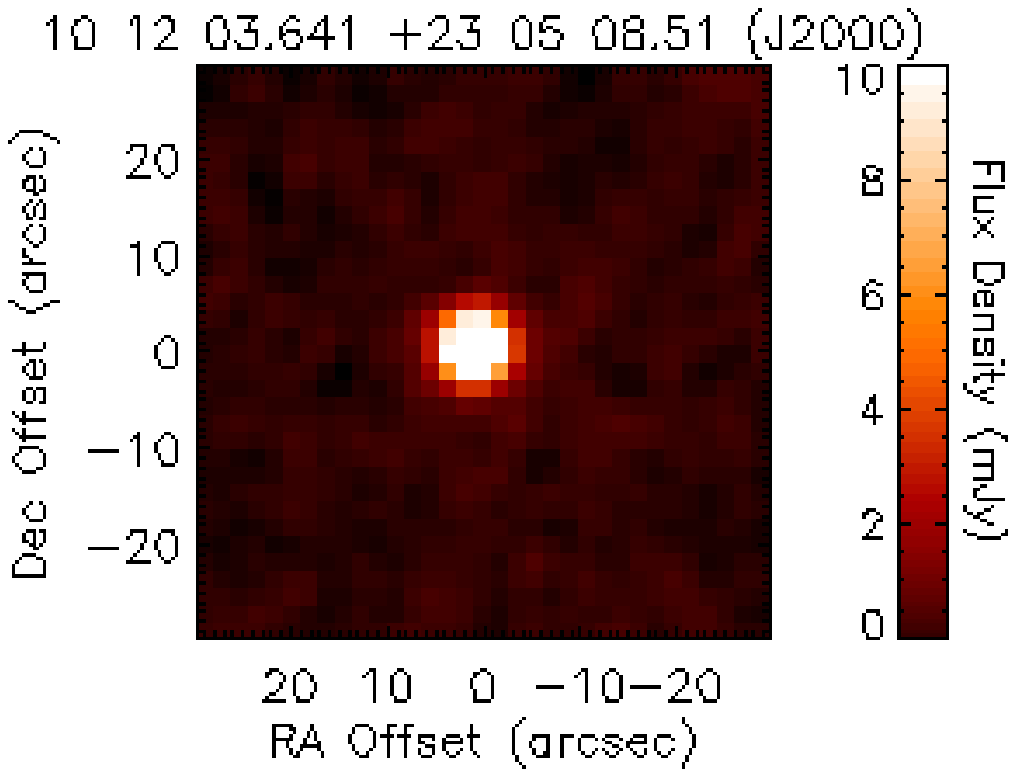}
\includegraphics[width=2.1in]{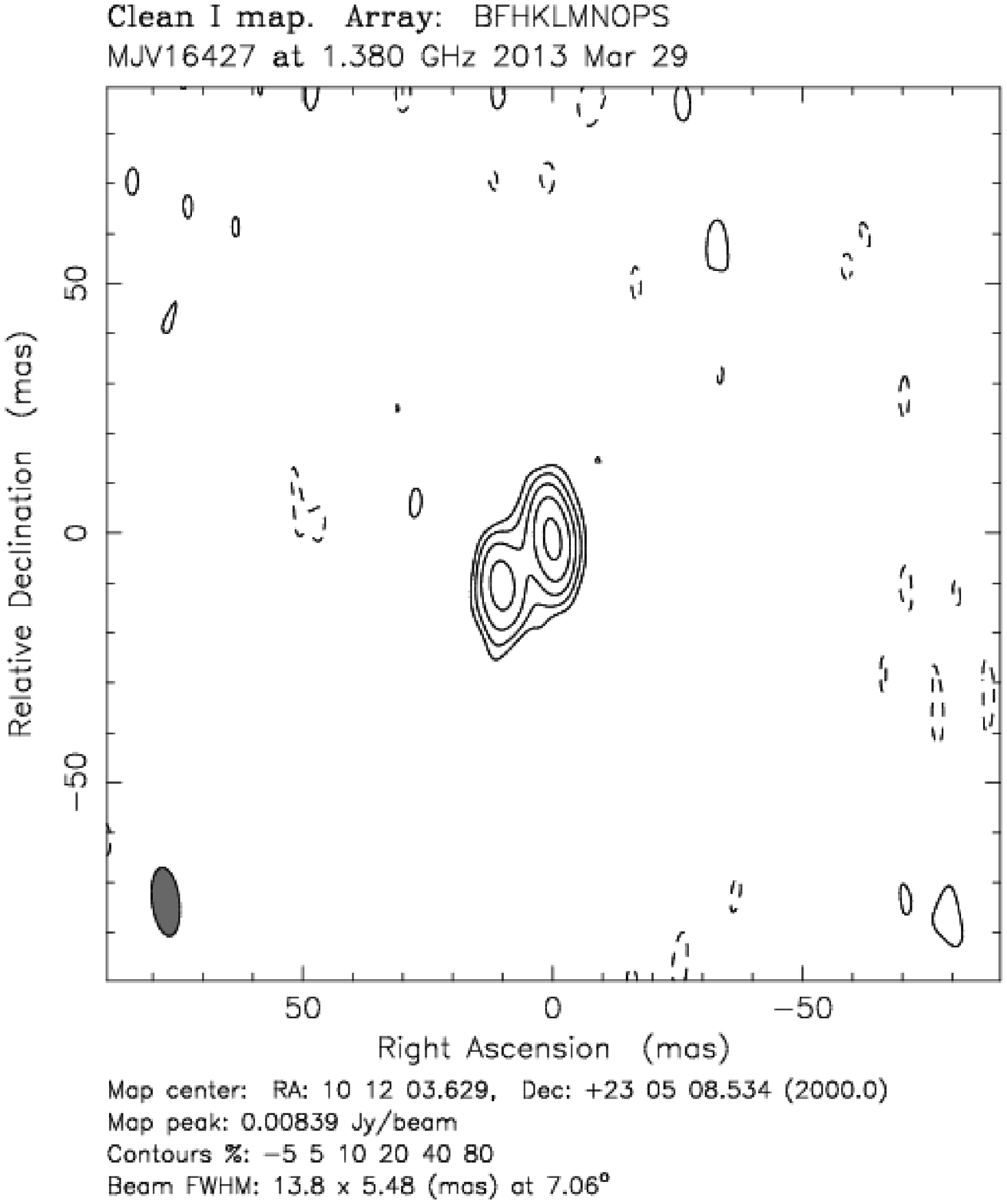}\\
\includegraphics[width=2.1in]{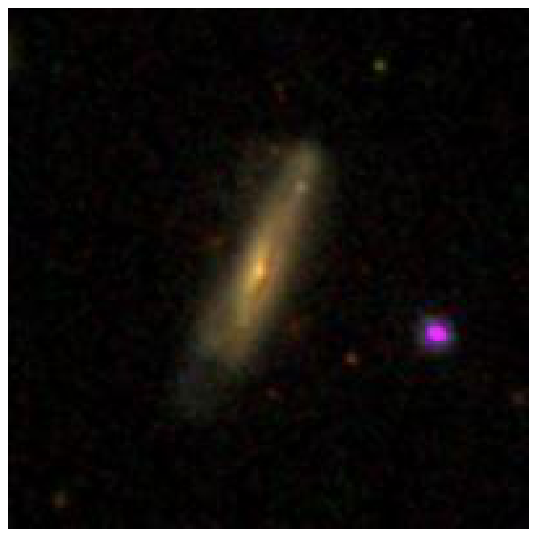}  & \includegraphics[width=2.1in]{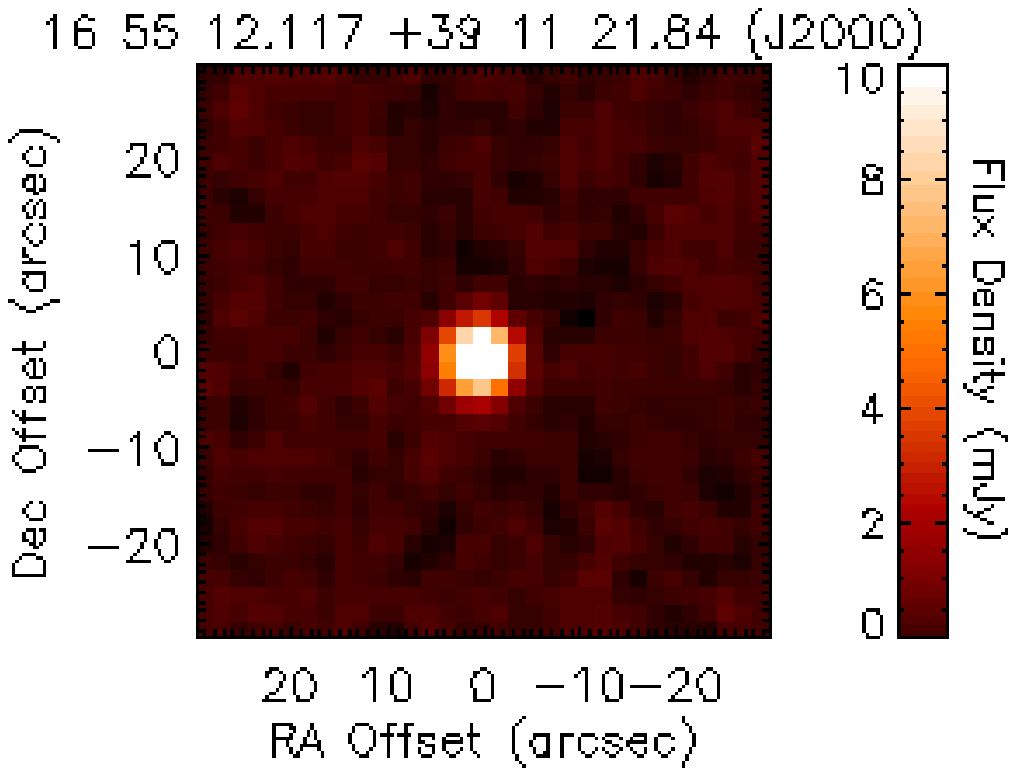}
\includegraphics[width=2.1in]{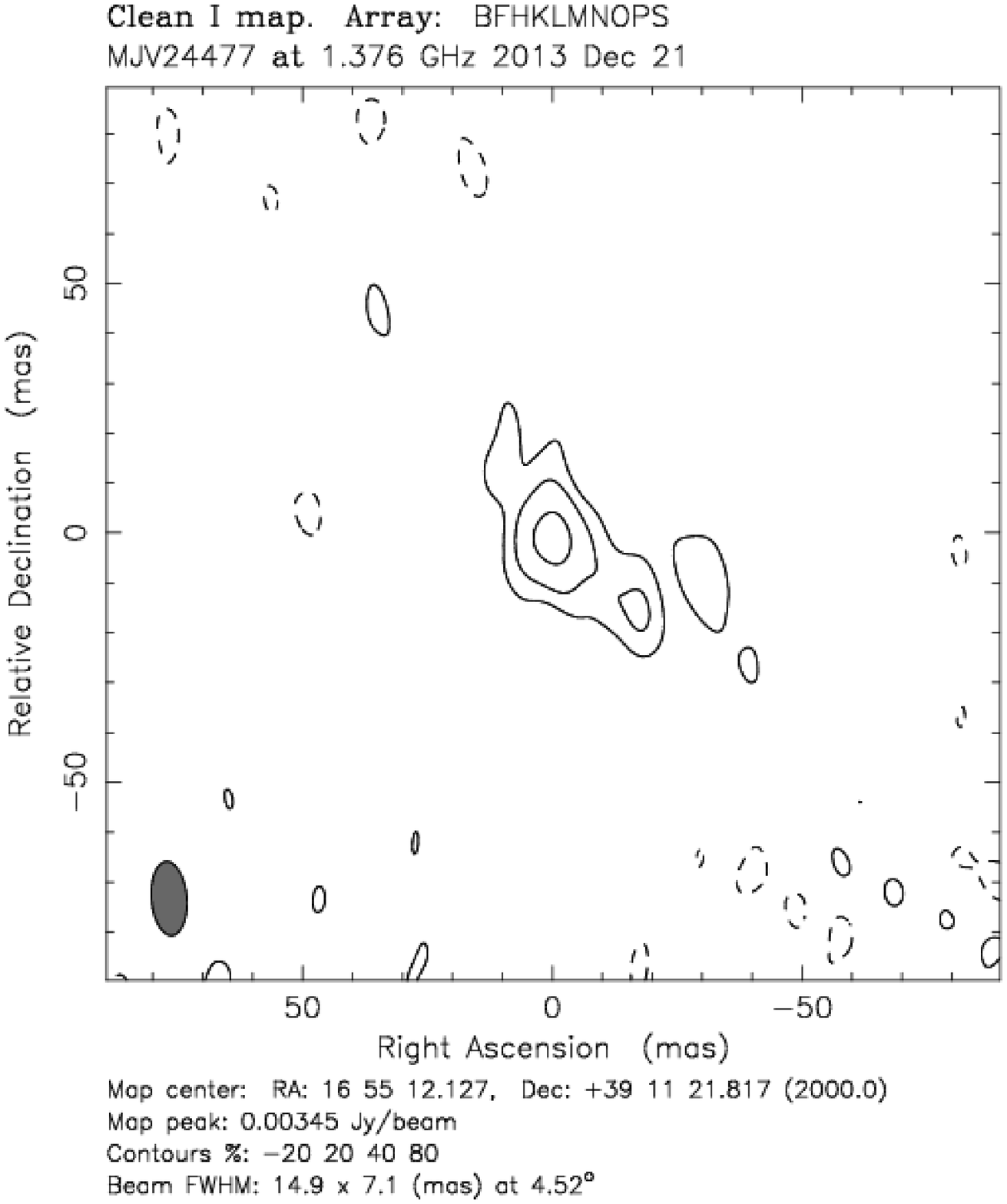}\\
\includegraphics[width=2.1in]{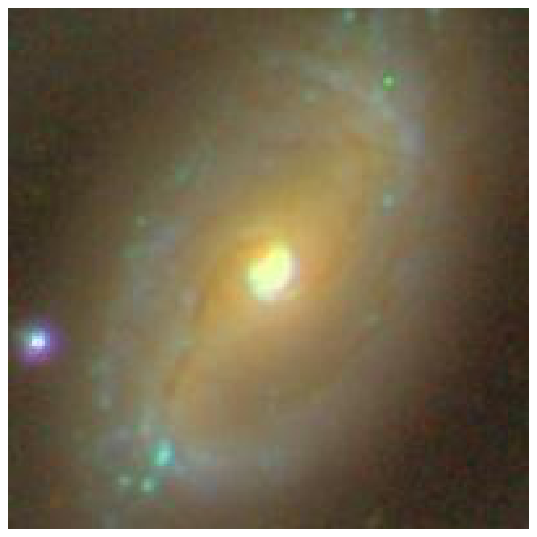}  & \includegraphics[width=2.1in]{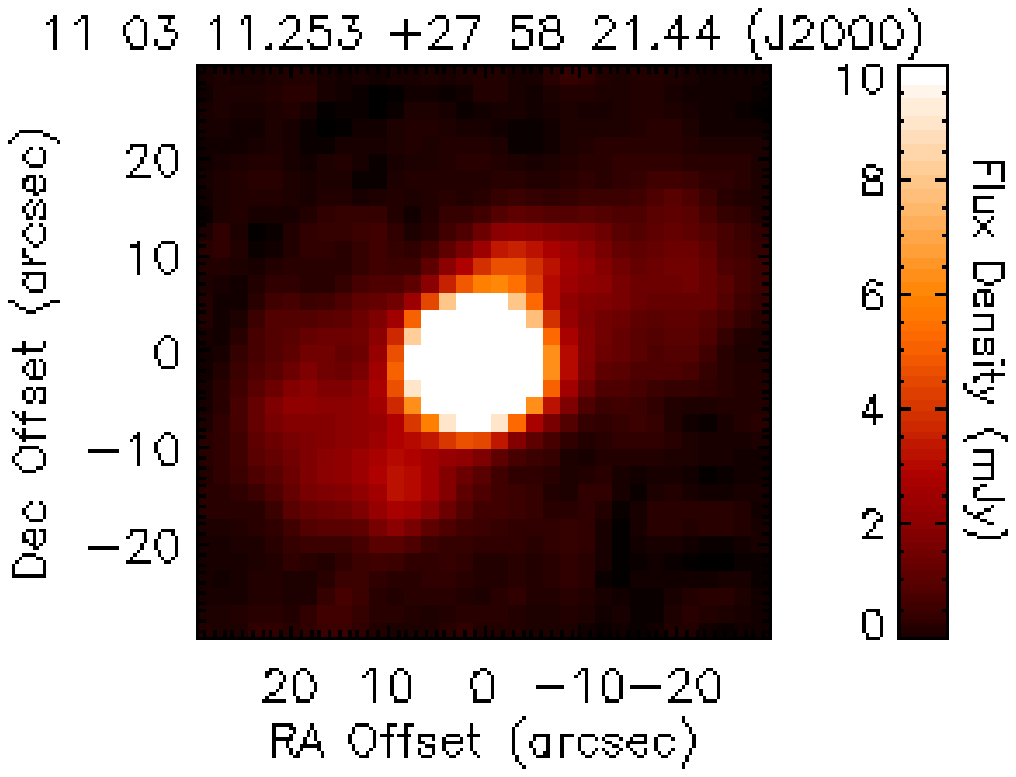}
\includegraphics[width=2.1in]{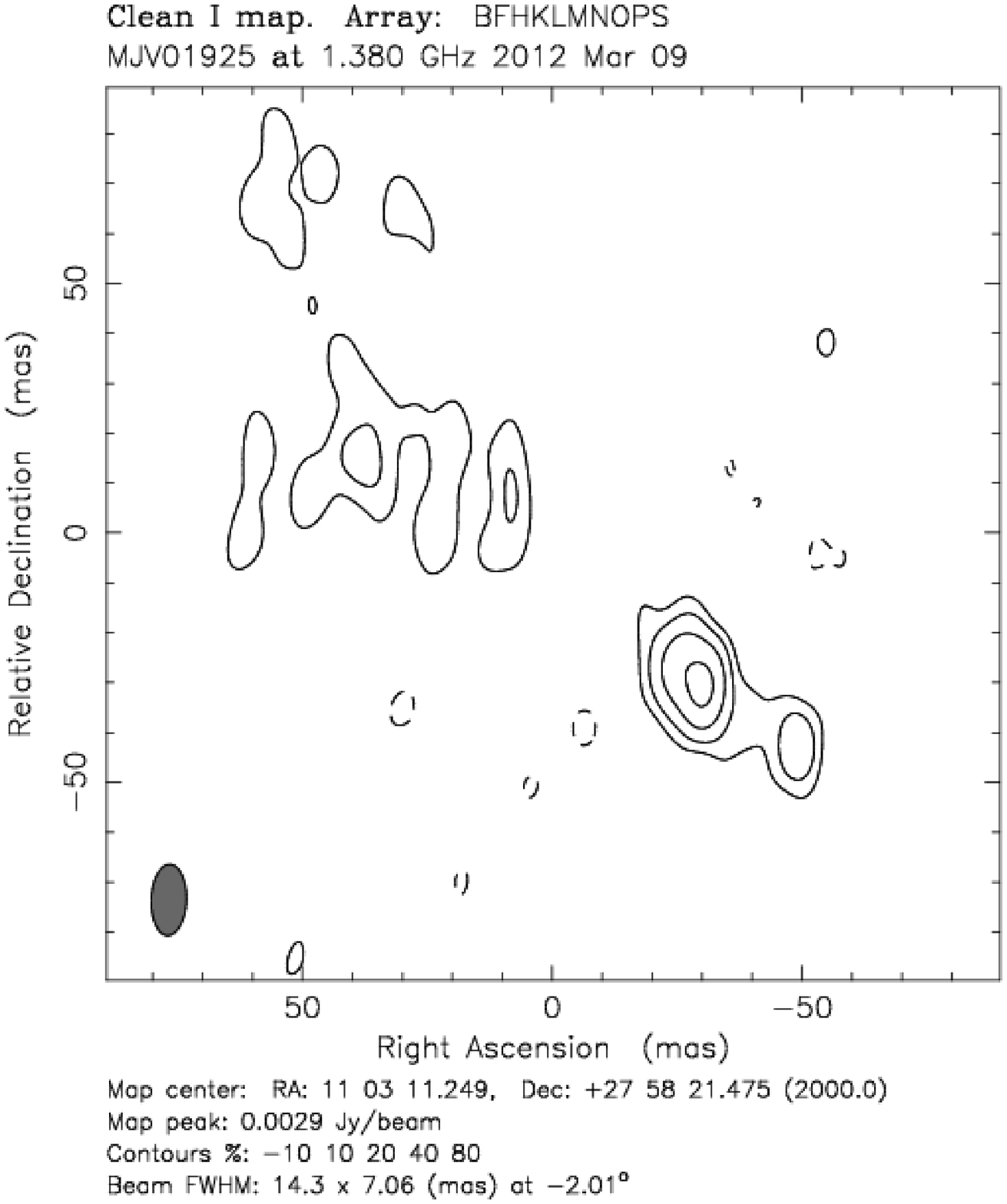}
\end{array}$
\caption{SDSS (left), FIRST (middle) and mJIVE-20 (right) images
of three spiral AGN hosts in our sample, that are close to edge-on
with evidence for a jet in the VLBI image. Note that, in the last
example (bottom row), the jet is shown slightly off centre, in
order to capture all of the structure around this galaxy on the
same spatial scales as for the other two
galaxies.}
\label{fig:VLBI_jets}
\end{center}
\end{minipage}
\end{figure*}

The analysis above indicates that local LTGs that host radio AGN
(with luminosities similar to those found in their early-type
counterparts) tend to be massive spirals with low SFRs, with
mergers being a plausible trigger for the onset of the AGN. It is
worth noting that that the requirements for LTGs to host radio AGN
are akin to those in their early-type counterparts that inhabit
similar (low-density) environments i.e. massive galaxies that have
red colours and are likely to be involved in a merger
\citep{Kaviraj2014a}. Nevertheless, the detection rate in
early-types is almost an order of magnitude higher
\citep{Kaviraj2014a}. While a thorough investigation is beyond the
scope of this particular paper, we speculate on some of the
potential reasons for this discrepancy. It is possible that hot
gas reservoirs in early-type galaxies could provide an extra
source of fuel for the black hole - we note, however, that the
\citet{Kaviraj2014a} sample does not contain central cluster
galaxies where hot gas cooling is expected to play a significant
role in AGN triggering
\citep[e.g.][]{Tabor1993,McNamara2007,Cattaneo2009,Fabian2012}.
Thus, it is unclear if the observed discrepancy in detection rate
can be completely explained solely via this argument.
Alternatively (or in addition), black hole masses may trace bulge
mass more closely than total galaxy stellar mass, implying that
even massive LTGs do not typically have black holes as massive as
those in their early-type counterparts. Finally, the `spin
paradigm', which postulates that radio jets are launched by
rapidly spinning black holes
\citep[e.g.][]{Blandford1977,Wilson1995,Sikora2007,Bagchi2014}
that are expected to reside primarily in bulge-dominated galaxies,
may also play a leading role in explaining the surfeit of radio
AGN in early-type systems compared to their late-type
counterparts.


\subsection{Implications for AGN feedback in spiral AGN hosts}
The recent literature indicates that, in low-density environments
in the \emph{nearby} Universe, AGN overwhelmingly avoid systems
that are in the blue cloud i.e. that are experiencing the peak of
their star-formation episode
\citep[e.g.][]{Schawinski2007,Wild2010,Carpineti2012,Kaviraj2011,Kaviraj2014b}.
Radio AGN in early-types, for example, cluster strongly on the red
sequence, where star formation rates have declined significantly
\citep{Kaviraj2014a}. Given that typical AGN lifetimes \citep[a
few times 10$^7$ yr, e.g.][]{Shabala2008} are much shorter than
the transit time from blue cloud to red sequence, this suggests a
time lag between the onset of star formation and the triggering of
AGN. Interestingly, this time lag shows no dependence on the
intensity of star formation (and therefore the gas fraction) in
the host system and is observed across the entire spectrum of star
formation activity, from intensely-starbursting luminous infrared
galaxies \citep{Kaviraj2009a} to the relatively weakly
star-forming early-type systems
\citep[e.g.][]{Wild2010,Shabala2012}. The implication of this time
delay is that the AGN typically does not couple to the host's gas
reservoir during the main starburst phase, thus reducing its
ability to strongly regulate the stellar mass growth.

Our study reinforces this picture, by extending these results to
spiral AGN hosts. In a similar vein to what is found in early-type
galaxies, radio AGN in the nearby LTG population preferentially
inhabit systems on the red sequence that have low SFRs. Unless the
typical AGN lifetime (i.e. the duration of the `on' phase) is
significantly longer in LTGs compared to their early-type
counterparts, the radio AGN in our spiral hosts are also likely to
have been triggered after the peak of the star formation activity
has elapsed in these systems, limiting their ability to impart
feedback on the gas reservoir. Taken together with the recent
literature, our results indicate that the time delay between the
onset of merger-driven star formation and the subsequent
triggering of AGN is likely to be a ubiquitous feature of radio
AGN in the nearby Universe.


\subsection{Jet orientations in spiral AGN hosts}
We conclude this section by exploring the morphology of the jets
in our VLBI-detected spirals. Three galaxies in our sample show
clear VLBI-scale jets. In Figure \ref{fig:VLBI_jets}, we present
the VLBI radio maps of these galaxies, together with their optical
SDSS and arcsecond-scale radio images from FIRST. In all cases,
the VLBI-scale jets are clearly close to being orthogonal to the
galaxy disc. Since the jet is expected to align with the spin axis
of the black hole \citep[e.g.][]{McKinney2013}, the orthogonality
observed in our galaxies is not unexpected, since co-evolution of
the black hole and the host spiral should produce alignment
between the black hole's spin axis and the stellar disc. In such a
scenario one would expect the jet to be orthogonal to the major
axis of the spiral galaxy, as is observed in our galaxies.

It is worth recalling that a high fraction (50\%+) of our LTG AGN
are likely merger remnants. One might expect that galaxy mergers
could disrupt the alignment between the black hole spin axis and
the stellar disc e.g. due to the coalescence of two similar mass
black holes in a major merger. However, it is important to note
that the interactions in question here are, by definition,
\emph{minor} mergers (i.e. mergers with small satellites), because
the discs in our LTG AGN remain intact, contrary to what might be
expected in major-merger remnants
\citep[e.g.][]{Barnes1992a,Springel2005a}. Our results therefore
suggest that minor mergers are unlikely to disrupt the alignment
between the black hole spin axis and the stellar disc (which may
naturally come about due to secular co-evolution of these two
systems)\footnote{Note that the well-known misalignment of
kpc-scale radio jets with the galaxy disk \citep{Ulvestad1984}
does not pose a serious problem to this interpretation, since
parsec and kpc-scale radio jets are often spectacularly
misaligned, as illustrated in Cen A \citep{Israel1998,Feain2009}.}

Finally, it is worth noting that several authors have studied the
relative geometry of galaxy-scale and circumnuclear structures in
systems that host AGN. Most recently, \citet{Greene2013} used
H$_2$O megamaser observations to map the circumnuclear discs in
seven nearby Seyfert galaxies, and found that while the maser
discs are oriented perpendicular to the direction of VLBI-scale
radio jets (consistent with jet generation models in which the
magnetic field associated with the black hole's accretion disk
determines jet orientation, e.g. \citet{McKinney2013}), the maser
and galaxy discs did not align in most of their sample. While the
reasons for this discrepancy with our results are difficult to
probe further using the available data, several causes for such
misalignments are possible, including disk warping on parsec
scales, subparsec-scale torques from stars in a cusp around the
black hole and natural changes in the angular momentum vector of
(secularly) infalling gas as it is torqued by the stars
\citep[e.g.][]{Maloney1996,Pringle1997,Hopkins2010,Gammie2000,Greene2013}.


\section{Summary and discussion}
We have combined the mJIVE-20 radio VLBI survey with the SDSS to
study the rare population of radio AGN in LTGs that have radio
luminosities similar to the more commonly-found early-type AGN in
the same stellar mass and redshift ranges. As noted in the
introduction, the study of radio AGN in LTGs is desirable, not
just because they are rare, but also because they are good
laboratories for exploring the behaviour of radio jets in disc
environments. While this is relatively rare at low redshift, the
connection between black holes and discs is likely to have been a
common feature of galaxy evolution around the peak epoch of star
formation ($z\sim2$) when most of today's stellar mass was
created. Radio AGN in LTGs may therefore hold clues to the
interplay between AGN and star formation at these critical epochs.
Our main results are as follows:

\begin{itemize}

    \item The radio AGN in LTGs are predominantly found in galaxies that have high stellar masses (M$_* > 10^{10.8}
    M_{\odot}$), red colours and low star formation rates, with a negligible dependence on the detailed morphology or local environment of the host
    galaxy.

    \item  A high fraction ($\sim50$\%) of spiral AGN hosts
    exhibit morphological disturbances that are indicative of recent mergers. The
    merger fraction is around a factor of 4 higher than that in
    the general LTG population, indicating that mergers are likely to be an
    important trigger for the radio AGN in these systems.

    \item The fraction of LTG AGN hosts that reside on the red sequence is
    almost a factor of 2 higher than that in the general LTG population. In a similar vein
    to what has been suggested in the recent literature, the predominantly red colours of the LTG AGN
    hosts suggest that the triggering of radio AGN in these systems
    is generally delayed with respect to the peak of the associated star formation
    episode, reducing the AGN's ability to regulate the stellar mass growth. Together with the recent
    literature, it seems reasonable to suggest that in starbursts
    where the the gas is accreted via a merger in the nearby Universe, AGN feedback is unlikely
    to strongly regulate the resultant star formation.

    \item In LTGs that exhibit parsec-scale jets in the VLBI images,
    we find that the orientations of these jets appear to be roughly perpendicular to the
    major axis of the host galaxy. If the black holes and discs
    co-evolve in these systems, then one might expect the black
    hole spin axes to align with the stellar discs which, in turn, would produce the observed alignment between the galaxy disc
    and the jet.

\end{itemize}


\section*{Acknowledgements}
We are grateful to Martin Hardcastle for many illuminating
discussions. We thank Meg Urry and Kevin Schawinski for comments.
SK is grateful for support from the University of Tasmania (UTas)
via a UTas Visiting Scholarship and acknowledges a Senior Research
Fellowship from Worcester College Oxford. SSS acknowledges an ARC
Early-Career Fellowship (DE130101399). ATD was supported by an NWO
Veni Fellowship.


\bibliographystyle{mn2e}
\bibliography{references}


\end{document}